\documentclass[11pt,twoside]{article}

\usepackage{asp2006}
\usepackage{epsf}
\usepackage{psfig}
\usepackage{lscape}
\usepackage{graphicx}

\begin{document}
\title{AMIGA project: Quantification of the isolation of 950 CIG galaxies}   
\author{S. Verley$^{1,2,3}$, S. Leon$^2$, L. Verdes-Montenegro$^2$, F. Combes$^3$, J.~Sabater$^2$, J. Sulentic$^2$, G. Bergond$^{2,4}$, D. Espada$^{2,5}$, U. Lisenfeld$^1$, S.~C.~Odewahn$^6$}   
\affil{$^1$Universidad de Granada, $^2$Instituto de Astrof\'isica de Andaluc\'ia -- CSIC, $^3$LERMA-OBSPM, $^4$CAHA, $^5$Harvard-Smithsonian Center for Astrophysics, $^6$Mc Donald Observatory}    

\begin{abstract} 
The role of the environment on galaxy evolution is still not fully understood. In order to quantify and set limits on the role of nurture one must identify and study a sample of isolated galaxies. The AMIGA project "Analysis of the Interstellar Medium of Isolated GAlaxies" is doing a multi-wavelength study of a large sample of isolated galaxies in order to examine their interstellar medium and star formation activity. We processed data for 950 galaxies from the Catalogue of Isolated Galaxies (CIG, Karachentseva 1973) and evaluated their isolation using an automated star-galaxy classification procedure (down to $M_B \sim 17.5$) on large digitised POSS-I fields surrounding each isolated galaxy (within a projected radius of at least 0.5~Mpc). We defined, compared and discussed various criteria to quantify the degree of isolation for these galaxies: e.g. Karachentseva's revised criterion, local surface density computations, estimation of the external tidal force affecting each isolated galaxy. We found galaxies violating Karachentseva's original criterion, and we defined various subsamples of galaxies according to their degree of isolation. Additionally, we sought for the redshifts of the primary and companion galaxies to access the radial dimension. We also applied our pipeline to triplets, compact groups and clusters and interpret the isolated galaxy population in light of these control samples.
\end{abstract}


\section{Introduction}

Although it is now generally recognised that the environment experienced by the galaxies during their whole lifetime plays a role as important as the initial conditions of their formation, there are still many open questions. In order to define what is the amplitude and dispersion of a given galaxy property that can be ascribed to ``nature'', a well characterised reference sample of isolated galaxies is needed. To be statistically useful the sample must be large enough to allow us to assess environmental effects both as a function of morphological type and luminosity. The motivation of the AMIGA ({\bf A}nalysis of the Interstellar {\bf M}edium of {\bf I}solated {\bf GA}laxies {\tt http://amiga.iaa.es}) project is to identify such a sample of isolated galaxies.

\section{The AMIGA project}

The AMIGA project adopted the Catalogue of Isolated Galaxies \citep[CIG:][]{1973AISAO...8....3K} as a starting point. The strength of the CIG involves its size (1050 galaxies) and its selection with a strong isolation criterion. Previous papers in this series included: 1) optical characterisation including derivation of the optical luminosity function \citep{2005A&A...436..443V}, 2) morphological revision using POSS-II (and SDSS overlap) and type-specific OLF analysis \citep{2006A&A...449..937S} and 3) mid- and far-infrared properties using the IRAS database \citep{2007A&A...462..507L}. Studies of the radio continuum \citep{2008A&A...485..475L}, \ion{H}{i} \citep{2005A&A...442..455E}, CO and H$\alpha$ emission \citep{2007A&A...474...43V} properties have also been carried along with a study of the small AGN population found in the sample \citep{2008A&A...486...73S}. This work focuses on a reassessment of the isolation degree for all galaxies in the CIG with recession velocities $V_{\rm r} \geq 1500$~km~s$^{-1}$.

\section{Catalogue of neighbours around isolated galaxies}

\begin{figure}
 \includegraphics[width=0.49\textwidth]{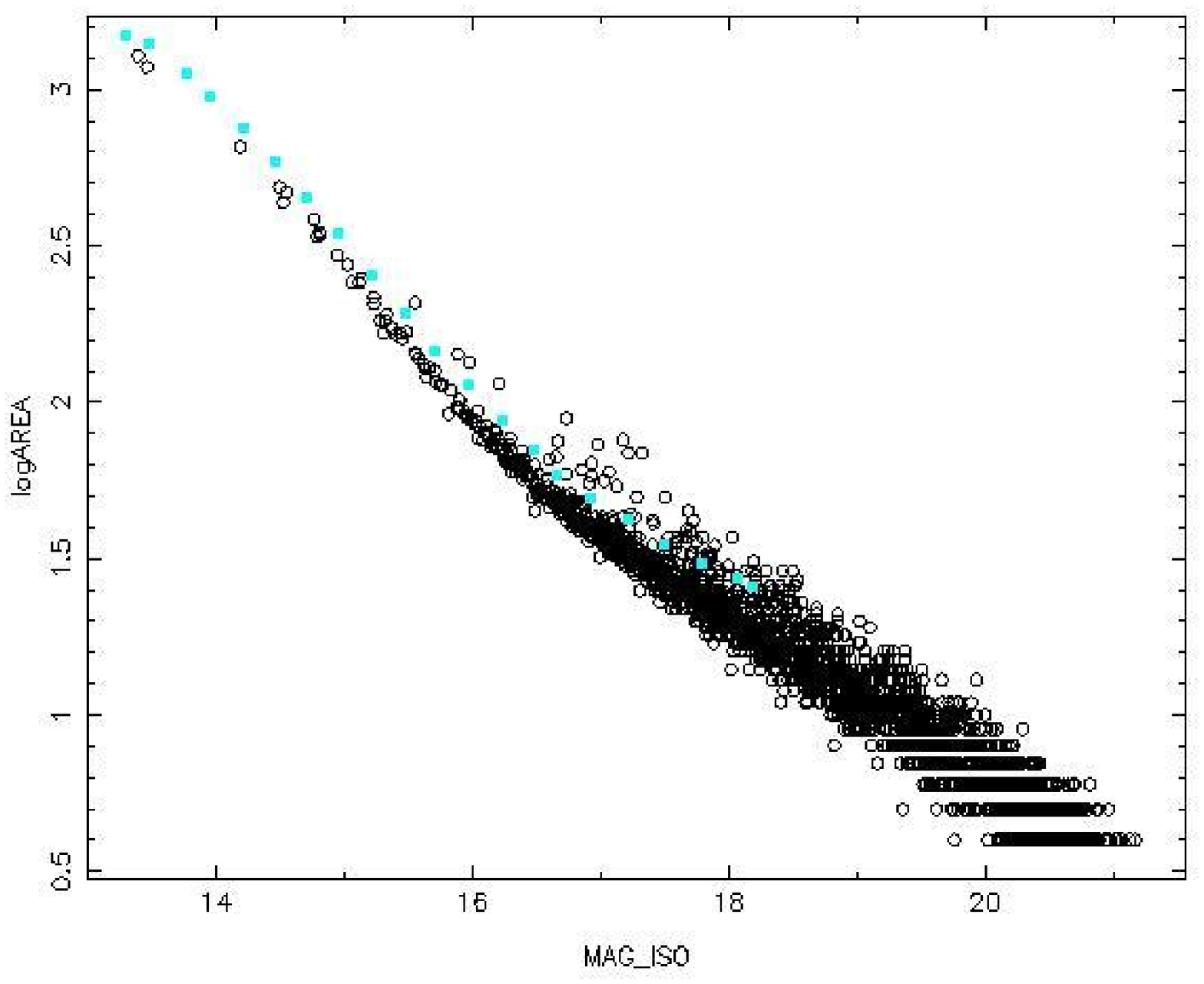}
 \includegraphics[width=0.49\textwidth]{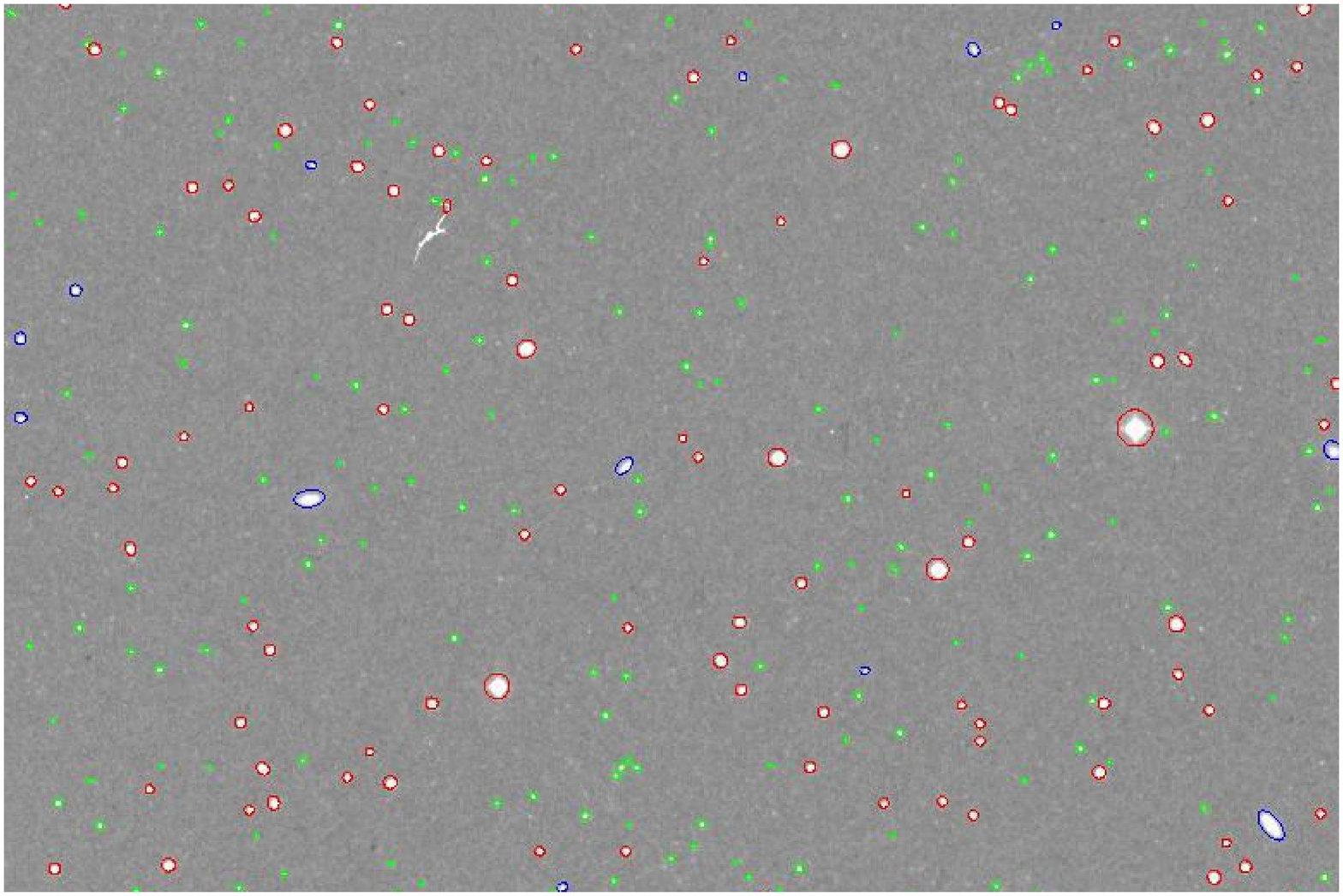}
\caption{{\it Left:} Star/galaxy separation parameter plane. The objects above the separation line (blue points) are classified as {\sc Galaxy}, while the ones below are classified as {\sc Star}. The objects fainter than the extent of the separation line are not classified, their type is {\sc Unknown}. {\it Right:} Close-up view of the distribution of galaxies around CIG 714 (the bottom-right galaxy). The detected galaxies are marked with blue ellipses, the stars are marked by red ellipses. The objects too small or to faint to be assigned a type are shown by the green ellipses. The plate defects were removed from our object extraction.} \label{fig:paramSpace714}
\end{figure}

By separating stars from galaxies (based on POSS-I plates), we were able to create a catalogue of more than 53\,000 neighbour galaxies, lying around the 950 CIG galaxies. The catalogues are available in electronic form at the CDS \citep{2007yCat..34700505V} or from {\tt http://amiga.iaa.es}. 
A typical star/galaxy separation parameter plane from a POSS-I E image (CIG 714) is shown in Fig.~\ref{fig:paramSpace714}. The galaxies have a lower surface brightness than the stars and in the $\log$(area) vs. magnitude plane, the two classes of objects fall in different loci.

\section{Quantification of the isolation}

\begin{figure}
 \includegraphics[width=\textwidth]{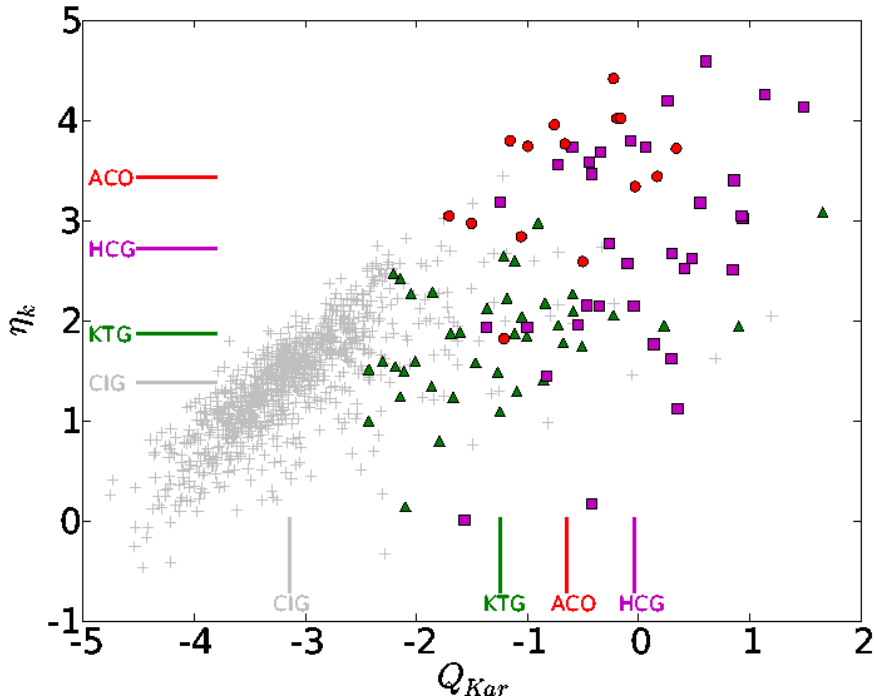}
\caption{Comparison between the local number density and tidal strength ($Q_{Kar}$) parameters for the CIG and the comparison samples. The CIG galaxies are represented by grey pluses. The KTG are depicted by green triangles, the HCG by magenta squares and the ACO by red dots. The mean values of each sample are shown by horizontal and vertical lines, following the same colour code. The catalogues are available at the CDS \citep{2007yCat..34720121V}.} \label{fig:cigKtgHcgAco}
\end{figure}

Using the $\sim$53\,000 neighbour galaxies, we calculated continuous parameters of isolation for 950 galaxies in the CIG. We used the local number density and the tidal force estimation to precisely describe the environment of each of the 950 CIG galaxies.

The local number density of the neighbour galaxies is calculated by focusing on the vicinity of the isolated galaxy candidates, where the principal perturbers should lie. An estimation of the local number density, $\eta_k$, is found by considering the distance to the $k^{\rm th}$ nearest neighbour. To probe a local region around the central galaxy, we consider $k$ equal to 5, or less if there are not enough neighbours in the field:
\[
\eta_k \propto \log\bigg(\frac{k-1}{V(r_k)}\bigg)
\]
with the volume $V(r_k) = 4 \pi r_k^3 /3$, where $r_k$ (in $'$) is the projected distance to the $k^{\rm th}$ nearest neighbour.

In order to provide an estimation of the degree of isolation taking also into account the masses of the neighbours, we calculated the tidal strength affecting the primary CIG galaxies. The tidal strength per unit mass produced by a neighbour is proportional to $M_i R_{ip}^{-3}$, where $M_i$ is the mass of the neighbour, and $R_{ip}$ is its distance from the centre of the primary ($\Delta R$ is the extension of the object). We approximated $R_{ip}$ by the projected separation, $S_{ip}$, at the distance of the CIG galaxy. Accordingly, the estimator $Q_{ip}$ is defined as the ratio between the tidal force and binding force:

\[Q_{ip} \equiv \frac{F_{\rm tidal}}{F_{\rm bind}} \propto \bigg(\frac{M_i}{M_p}\bigg) \bigg(\frac{D_p}{S_{ip}}\bigg)^3 \propto \frac{(\sqrt{D_p D_i})^3}{S_{ip}^3}\]

The logarithm of the sum of the tidal strength created by all the neighbours in the field, $Q = \log(\sum_i Q_{ip})$, is a dimensionless estimation of the gravitational interaction strength.

Figure~\ref{fig:cigKtgHcgAco} shows the comparison between the local number density and the tidal strength estimation for the CIG and the other catalogues sampling denser environments. These two isolation parameters are complementary and allow us to have a clear picture of the environment of each CIG galaxy considered. We compared the level of isolation of the galaxies in the CIG with galaxies in denser environments: 41 triplets \citep[KTG][]{1979AISAO..11....3K}, 34 compact groups \citep[HCG][]{1982ApJ...255..382H} and 15 clusters \citep[ACO][]{1958ApJS....3..211A} were selected for the comparison. The two isolation parameters are very well suited to discriminate the isolated galaxies from galaxies more affected by their environments. A more complete information can be found principally in \citet{2007A&A...470..505V} and \citet{2007A&A...472..121V}.

\end{document}